\documentclass[english,a4paper,12pt]{article}
\usepackage[T2A]{fontenc}
\usepackage[cp1251]{inputenc}
\usepackage{babel}
\usepackage{amsmath}
\usepackage{graphicx}
\usepackage{amssymb}
\usepackage{epsf}
\usepackage{pazh_eng}
\usepackage[labelsep=period,hang]{caption} %заменить умолчальное разделение ':' на '.'
\tightenlines

\newcommand{\WI}[2]{#1_{\mathrm{#2}}}
\sloppypar

\begin{document}
	\baselineskip 21pt
	
	%хдл 524.354.4
	
	\title{\bf Structure of Relativistic Stars Composed of Incompressible Matter
in the Absence of Strict Electroneutrality}
	
	\author{\bf \hspace{-1.3cm} \ \
		N.I. Kramarev\affilmark{*1,2}, A.V. Yudin\affilmark{2}}
	
	\affil{
		{\it Moscow State University, Moscow, 119991 Russia}$^1$\\
		{\it National Research Center ``Kurchatov Institute'' --- ITEP, Moscow, 117218 Russia}$^2$\\
%		{\it НИЦ ``Курчатовский Институт'', Москва, пл. Курчатова, 1}$^3$
}
	
	\vspace{2mm}
	%\received{~~~~~~~~}
%	\received{\today}
	%\revised{}
	
	\sloppypar
	\vspace{2mm}
	\noindent
The structure of a star composed of locally non-electroneutral incompressible three-component
matter is considered within the framework of general relativity. For thermodynamic quantities like the
pressure, the solution can be represented as a series in the small parameter $1/\WI{\Lambda}{G}\sim 10^{-36}$, where the
first approximation is the well-known electroneutral solution. However, the equilibrium equations for
the chemical potentials of the matter components, as it turns out, contain finite contributions from non-electroneutrality
effects even in the zeroth order. Analytical solutions have been obtained for all of the
parameters of the problem under consideration, which are illustrated by numerical examples.

	\noindent
	{\bf Keywords:\/} neutron stars, general relativity, stellar structure, electroneutrality of matter
	
	\noindent
	{\bf PACS codes:\/} 04.20.Jb, 95.30.Sf, 97.60.Jd
	
	\vfill
	\noindent\rule{8cm}{1pt}\\
	{$^*$ email: $<$kramarev-nikita@mail.ru$>$}
	
	\clearpage
	
	%***************************************************************
	\section*{INTRODUCTION}
	\noindent To calculate the stellar plasma parameters, the
local electroneutrality (LEN) approximation is commonly
used, i.e., the densities of positive and negative
charges are assumed to be strictly equal at each point.
This is explained by the extreme weakness of gravity
compared to other forces: for example, the ratio of the
force of electrostatic repulsion between two protons
to the force of their gravitational attraction is characterized
by the parameter
	\begin{equation}
	\WI{\Lambda}{G}=\frac{\mathbf{e}^2}{G\WI{m}{u}^2}\approx 1.25 \times 10^{36}.\label{lambda_G}
	\end{equation}
	Meanwhile, as was first shown by Rosseland (1924),
the plasma inside ordinary stars is polarized in their
gravitational field. There arises a constant local
charge imbalance characterized by the small parameter
$1/\WI{\Lambda}{G}$:
	\begin{equation}
	\frac{|\Delta n|}{n}\sim 10^{-36},
	\end{equation}
where $n$ is the local number density of matter and
 $\Delta n$ is the difference of the densities of positive and
negative charges. As a consequence, a large-scale
polarization field arises, and, in fact, each ion (or a
positively charged nucleus) is in equilibrium of two
forces: the gravitational field and the electrostatic
polarization field. The polarization problem as applied
to white dwarfs was considered, for example, in the
book by Schatzman (1958). In view of its weakness,
this field has virtually no effect on the stellar structure
and is taken into account only when calculating
the diffusion processes (see, e.g., Beznogov and
Yakovlev 2013; Gorshkov and Baturin 2008) in stars.
Thus, the LEN of matter is an excellent approximation
for calculating the structure and properties of
stars.

The structure of stars in the absence of LEN was
considered in a number of papers (see, e.g., Bally
and Harrison 1978; Neslusan 2001; Iosilevskiy 2009).
Krivoruchenko et al. (2018) explored the topic under
consideration in the Newtonian approximation using
two-component polytropic stellar models. The complete
solution determining the stellar structure turned
out to consist of two parts: a regular part that can be
represented as a series in the small parameter $1/\WI{\Lambda}{G}$
with the LEN solution as the zeroth approximation
and an irregular part that is exponentially small everywhere,
except for a finite number of zones usually
located at the boundary of the domain of integration
(the so-called boundary layer). This is because the
stellar equilibrium equations in the absence of LEN
refer to the so-called singularly perturbed problems
(see, e.g., O’Malley 1991), i.e., for the case where
a small parameter appears at the highest derivative
of the differential equation. However, the deviations
from LEN are small even in the complete solution
obtained, LEN is strongly violated only in a thin surface
layer (the so-called ``electrosphere''), the polarization
field is everywhere small, and the total stellar
charge can change only in a narrow range, $-0.1\lesssim Q\lesssim 150$~C.
	
However, all of what has been said above concerns
ordinary, nondegenerate stars or white dwarfs. In
their relatively recent papers, Rotondo et al. (2011)
and Belvedere et al. (2012, 2015) considered the effects
of deviation from LEN in neutron stars (NSs).
The authors assert that they obtained a solution in
which the proton density at the boundary of the NS
core is much greater than the electron density and
reaches a maximum there. This leads to a growth of
the electric polarization field, which reaches several
thousand Schwinger critical fields! This core is overlain
by an electrosphere of electrons compensating
for the large positive charge of the core, and a crust
that is a lattice of neutron-rich nuclei in a Fermi
sea of electrons rests on it (here, the authors used
the equation of state from Baym et al. (1971)). The
absence of an extended core–crust transition region,
from homogeneous nuclear matter to neutron-rich
nuclei (see Fig.~17 from Belvedere et al. 2012), leads
to deviations, in particular, of such macroscopic parameters
as the NS mass and radius from the values
predicted by the classical LEN solution (for the latter,
see, e.g., Haensel et al. 2007; Pearson et al. 2018). In
view of such discrepancies, we realized the need for a
further study of the effects of deviations from LEN in
degenerate stars.
	
This paper is the first step in this direction. Using
the previous experience (Krivoruchenko et al. 2018)
and the fundamental papers by Olson and Bailyn
(1975, 1978), who derived the equilibrium equations
for matter within the framework of general relativity
in the absence of LEN, we managed to generalize
the past calculations to the case of a multi-component
fluid in general relativity. So far we have restricted
our analysis to incompressible nuclear matter (the
polytrope $n=0$). Here, we did not consider the
above-mentioned problems associated with the NS
crust either. Despite being artificial, this simple (toy)
model allows one not only to ``feel the physics'' of
the problem within the framework of general relativity,
but also to obtain important results. For
example, a constraint on the NS mass $M$ and radius
$R$ was previously deduced in this approximation,
$GM/Rc^2<4/9$, which also remains valid in the
general case (Weinberg 1972). It is also important
that the polytrope n = 0 has no irregular component
of the solution (see the Appendix in Krivoruchenko
et al. (2018)), which simplifies it considerably. In this
paper, we even managed to obtain an analytical solution
for the model under study within the framework
of general relativity.
	
The paper is organized as follows: first, we write
the basic equations of the problem in general form.
Then, using the approximation of incompressible
matter, we simplify the equations and bring them to
dimensionless form. The solutions obtained are then
illustrated with several examples. Next, we briefly
discuss the electrospheres in this approximation and
provide our conclusions.

	%*************************************************************
	\section*{BASIC EQUATIONS}
	\noindent Let us write the equations of the problem in the
form given in Olson and Bailyn (1978):
	\begin{gather}
	\frac{dm}{dr}=4\pi r^2\left[\rho+\frac{Q^2}{8\pi c^2r^4}\right],\label{dmdr}\\
	\frac{d Q}{dr}=4\pi r^2 e^{\lambda/2}\sum q_i n_i,\label{dEdr}\\
	\frac{d\mu_k}{dr}=e^{\lambda/2}\frac{q_k Q}{r^2}-e^{\lambda}\frac{G\mu_k}{r^2 c^2}
	\left[m-\frac{Q^2}{2 c^2 r}+\frac{4\pi r^3}{c^2}\left(\sum n_i\mu_i{-}\rho c^2\right)\right].\label{dmudr}
	\end{gather}
Here, Eq. (\ref{dmdr}) is the continuity equation, $r$ is the radial
coordinate, $m$ is the mass coordinate, and $\rho$ is the
mass–energy density. The quantity $Q$ denotes the
total charge within a sphere of radius $r$ defined by
Eq. (\ref{dEdr}), where the sum on the right-hand side is
over all matter components with charges $q_i$, in our
case, $i=\mathrm{n}, \mathrm{p}, \mathrm{e}$. The metric function $\lambda$ is defined in
a standard way:
	\begin{equation}
	e^{\lambda}=-\WI{g}{11}=\left[1{-}\frac{2 G m}{r c^2}\right]^{-1},\label{lambda}
	\end{equation}
where $\WI{g}{11}$ is the corresponding component of the
Schwarzschild metric (see, e.g., Landau and Lifshitz
1975). Equations (\ref{dmudr}) are the equilibrium equations
for the chemical potentials $\mu_k$ of the matter
components. We will use the thermodynamic relations
(assuming that the temperature is $T = 0$)
	\begin{gather}
	\Phi\equiv\sum\mu_i n_i=\rho c^2+P,\label{thermod1}\\
	dP=\sum n_i d\mu_i,\label{thermod2}
	\end{gather}
where $P$ is the matter pressure. Multiplying each
of Eqs. (\ref{dmudr}) by $n_k$ and adding them, we will obtain
the equilibrium equation for stellar matter (Olson and
Bailyn 1975):
	\begin{equation}
	\frac{dP}{dr}=\frac{Q}{r^2}\frac{\sum q_i n_i }{\sqrt{1-\frac{2Gm}{rc^2}}}-\frac{G\left(\rho+\frac{P}{c^2}\right)}{r\left(r-\frac{2Gm}{c^2}\right)}
	\left(m+\frac{4\pi r^3}{c^2}P-\frac{Q^2}{2rc^2}\right),\label{TOV}
	\end{equation}
In the case of strict electroneutrality $(Q = 0)$, it
turns into the Tol\-man\--Op\-pen\-hei\-mer\--Vol\-koff (TOV)
equation.
	
	%*************************************************************
	\section*{POLYROPE $n = 0$}
	\noindent In what follows, we will restrict ourselves to the
absolutely stiff equation of state in which the number
densities of the matter components $n_k$ are constant.
It corresponds to the polytrope $n = 0$ in the non-relativistic
case. Since the internal energy $\epsilon$ of the
polytrope matter is related to the pressure by the
relation $\epsilon=nP/\rho$, the total mass–energy density $\rho$
is $\rho=\sum n_i m_i$. It can then be seen from Eq. (\ref{thermod1})
that in the case under consideration the thermodynamic
potential $\Phi$ differs from the pressure $P$ by a
constant. Let us use this and rewrite the equilibrium
equation (\ref{TOV}) as
	\begin{equation}
	\frac{d\Phi}{dr}=\frac{1}{8\pi r^4}\left(\frac{d Q^2}{dr}\right)+e^{\lambda}\frac{G\Phi }{c^2}\left[\frac{d}{dr}\left(\frac{m}{r}\right)-\frac{4\pi r}{c^2}\Phi\right],\label{eqPhiv1}
	\end{equation}
where we also used Eqs. (\ref{dmdr}) and (\ref{dEdr}). Equation (\ref{eqPhiv1})
can be written in an equivalent, more convenient
form:
	\begin{equation}
	\frac{d}{dr}\left(\Phi e^{{-}\lambda/2}\right)=\frac{e^{{-}\lambda/2}}{8\pi r^4}\left(\frac{d Q^2}{dr}\right)-\frac{4\pi G r}{c^4}\Phi^2 e^{\lambda/2}.\label{eqPhiv2}
	\end{equation}
	The solution of the homogeneous equation (\ref{eqPhiv2}) is:
	\begin{equation}
	\Phi e^{{-}\lambda/2}=\left[C+\int\frac{4\pi G r}{c^4}e^{3\lambda/2} dr\right]^{{-}1},\label{PHI0}
	\end{equation}
where $C$ is some constant. We seek a solution of the
complete equation (\ref{eqPhiv2}) by the method of variation of
constants. For $C = C(r)$ we then obtain the differential
equation:
	\begin{equation}
	\frac{dC(r)}{dr}=-\frac{e^{{-}\lambda/2}}{8\pi r^4}\left(\frac{d Q^2}{dr}\right)\left[C(r)+\int\frac{4\pi G r }{c^4}e^{3\lambda/2}dr\right]^2.\label{dCdr}
	\end{equation}
Let us now return to the equilibrium equations (\ref{dmudr}).
They can be rewritten as	
	\begin{equation}
	\frac{d\mu_k}{dr}=\frac{q_k Q}{r^2}e^{\lambda/2}+e^{\lambda}\frac{G\mu_k}{c^2}\left[\frac{d}{dr}\left(\frac{m}{r}\right)-\frac{4\pi r}{c^2}\Phi\right] \label{dmudr_new}.
	\end{equation}
	The solution of these equations can be written as
	\begin{equation}
	\mu_k(r) e^{{-}\lambda(r)/2}=e^{{-}\Pi(r)}\Big[\mu_k(0)+\int\limits_0^r\frac{q_k Q(r')}{r'^2}
	e^{\Pi(r')}dr'\Big],\label{mu_equil}
	\end{equation}
	where we introduced the notation
	\begin{equation}
	\Pi(r)\equiv\int\limits_0^r\frac{4\pi G r'}{c^4}e^{\lambda(r')}\Phi(r')dr'.\label{Pi_def}
	\end{equation}
	
	\subsection*{Dimensionless Form of the Equations}
	\noindent Before proceeding to the solution of the derived
equations, let us make several simplifying assumptions.
First, we will assume that the matter is in
beta equilibrium, i.e., $\WI{\mu}{n}=\WI{\mu}{p}+\WI{\mu}{e}$. As can be seen
from Eq. (\ref{mu_equil}), if this condition is fulfilled at least at
one point of the star, then it is fulfilled everywhere
(because $\WI{q}{n}=0$ and $\WI{q}{p}=-\WI{q}{e}$). Second, we will seek
a solution in which both total and partial pressures
of the components \emph{become zero at one point}. At
this point, the chemical potentials of the components
are $\mu_k=m_k c^2$. Some contradiction with the
beta-equilibrium condition arises here, because $\WI{m}{n}\neq\WI{m}{p}+\WI{m}{e}$. 
To achieve consistency, we will assume that
$\WI{m}{e}=0$ and $\WI{m}{n}=\WI{m}{p}=\WI{m}{u}$, where $\WI{m}{u}$ is the atomic
mass unit. Such an approximation corresponds well
to the model nature of the problem being solved, in
particular, to the condition of absolute matter stiffness
and the constancy of the component number densities
following from it. In reality, the beta-equilibrium
condition will be violated in the outer stellar regions
as the density drops, and it is definitely violated in the
stellar electrosphere (see below and Krivoruchenko
et al. 2018).

Under the above assumptions, the matter density
is $\rho=\WI{\rho}{b}=\WI{m}{u}\WI{n}{b}$, where $\WI{n}{b}=\WI{n}{n}+\WI{n}{p}$ is the baryon
number density. We can now introduce a unit of
length natural for our problem:	
	\begin{equation}
	r_0=\frac{c}{\sqrt{4\pi G\WI{\rho}{b}}}\approx\frac{32.7}{\sqrt{\rho_{14}}}~\mbox{km},
	\end{equation}
where $\rho_{14}\equiv\rho(\mbox{g}/\mbox{cm}^3)\times 10^{-14}$. Let us introduce
the spatial coordinate $x$ according to $r=r_0 x$. The
mass coordinate $m$ is then expressed via the dimensionless
variable $\eta$  as $m(r)=4\pi r_0^3\WI{\rho}{b}\eta(x)$. Let us
also introduce the dimensionless charge $\theta$ according
to $Q(r)=4\pi r_0^3 \mathbf{e}(\WI{n}{p}{-}\WI{n}{e})\theta(x)$, where $\mathbf{e}$ is the unit of
electric charge. The stellar equilibrium equations (\ref{dmdr})
and (\ref{dEdr}) will then be written as follows:
	\begin{align}
	\frac{d\eta}{dx}&=x^2+\WI{\Lambda}{q}\frac{\theta^2}{x^2},\label{dmdx_dim}\\
	\frac{d\theta}{dx}&=x^2 e^{\lambda/2},\label{dEdx_dim}
	\end{align}
where, as above, $e^{{-}\lambda}=1-2\eta/x$ and the parameter
$\WI{\Lambda}{q}$ is
	\begin{equation}
	\WI{\Lambda}{q}=\frac{1}{2}\WI{\Lambda}{G}\left(\frac{\WI{n}{p}{-}\WI{n}{e}}{\WI{n}{b}}\right)^2.
	\end{equation}
Recall that the quantity $\WI{\Lambda}{G}\sim 10^{36}$ in this formula is
the main and, at the same time, huge parameter of the
problem.
	
Introducing the dimensionless variable $\phi$ for the
thermodynamic potential $\Phi$ according to $\Phi(r)=\WI{m}{u}c^2\WI{n}{b}\phi(x)$, we will rewrite relation (\ref{PHI0}) as
	\begin{equation}
	\phi(x) e^{{-}\lambda/2}=\left[\mathbf{c}(x)+\int e^{3\lambda/2} x dx\right]^{{-}1}.\label{Phi_dim}
	\end{equation}
Equation (\ref{dCdr}) will be written in dimensionless formas
follows:
	\begin{equation}
	\frac{d\mathbf{c}}{dx}=-\WI{\Lambda}{q}\frac{e^{{-}\lambda/2}}{x^4}\left(\frac{d\theta^2}{dx}\right)\left[\mathbf{c}(x)+\int e^{3\lambda/2}xdx\right]^2.\label{dCdr_dim}
	\end{equation}
Now it remains only to write the equilibrium equations
for the chemical potentials (\ref{mu_equil}) using the dimensionless
variables $\psi_k(x)\equiv\frac{\mu_k(r)}{\WI{m}{u}c^2}$:
	\begin{equation}
	\psi_k e^{{-}\lambda/2}=e^{{-}\Pi(x)}\Big[\psi_k(0)+\WI{\Lambda}{G}\!\left(\!\frac{\WI{n}{p}{-}\WI{n}{e}}{\WI{n}{b}}\!\right)\!\int\limits_0^x\frac{\tilde{q}_k \theta(x')}{x'^2}
	e^{\Pi(x')}dx'\Big],\label{mu_equil_dim}
	\end{equation}
where $\tilde{q}_k=0,\pm 1$ is the normalized dimensionless
charge of the neutron, proton, or electron, and $\Pi(x)$
defined by Eq. (\ref{Pi_def}) is
	\begin{equation}
	\Pi(x)\equiv\int\limits_0^x e^{\lambda(x')}\phi(x')x'dx'.\label{Pi_dim}
	\end{equation}
It is important to note that the huge factor $\WI{\Lambda}{G}\sim 10^{36}$
multiplied by the relative deviation of the matter number
densities from the electroneutral values appears
before the integral term in (\ref{mu_equil_dim}). This means that
the solution with dimensionless chemical potentials
of the order of $O(1)$ can be obtained only if
	\begin{equation}
	\frac{\WI{n}{p}{-}\WI{n}{e}}{\WI{n}{b}}=\frac{\WI{\alpha}{e}}{\WI{\Lambda}{G}},
	\end{equation}
where $\WI{\alpha}{e}=O(1)$ is a numerical parameter of the
problem. Then,
	\begin{equation}
	\WI{\Lambda}{q}=\frac{\WI{\alpha}{e}^2}{2\WI{\Lambda}{G}}\sim 10^{-36},
	\end{equation}
is a small parameter. The terms with the polarization
field enter into Eqs. (\ref{dmdx_dim}) and (\ref{dCdr_dim}) with this extremely
small factor. The equilibrium equations (\ref{mu_equil_dim}),
where the huge factor $\WI{\Lambda}{G}$ is compensated for by the
correspondingly small deviation of the matter from
electoneutrality, are an exception. This closely corresponds
to the case of Newtonian gravity considered
previously (Krivoruchenko et al. 2018; Hund
and Kiessling 2021a): the polarization field may be
neglected when calculating the stellar structure, but
the force with which this field acts on a separate
charged particle is comparable to the corresponding
gravitational force. Indeed, the stellar equilibrium
equation (\ref{TOV}) written in dimensionless variables ($p\equiv P/\rho c^2$) is
	\begin{equation}
	\frac{dp}{dx}=2\WI{\Lambda}{q}e^{\lambda/2}\frac{\theta}{x^2}-e^{\lambda}\frac{1{+}p}{x^2}
	\Big(\eta+x^3 p-\WI{\Lambda}{q}\frac{\theta^2}{x}\Big),\label{TOV_dim}
	\end{equation}
and all of the corrections to the standard TOV equation
(the terms with $\WI{\Lambda}{q}$) turn out to be negligible.

	\subsection*{Solution of the Equations}
	\noindent Given that the parameter $\WI{\Lambda}{q}$ is extremely small, a
solution can be sought in the form of a series in it. Let
us first turn to Eqs. (\ref{dmdx_dim}) and (\ref{dEdx_dim}). We will solve them
by the method of successive approximations. The
zeroth approximation for (\ref{dmdx_dim}) gives $\eta=\frac{x^3}{3}$. Then,
$e^{{-}\lambda}=1{-}\frac{2x^2}{3}$. Substituting this into (\ref{dEdx_dim}), we will
obtain the expression for the stellar charge in the first
approximation:
	\begin{equation}
	\theta(x)=\frac{3}{4}\left[\sqrt{\frac{3}{2}}\arcsin\sqrt{\frac{2}{3}}x-x\sqrt{1-\frac{2x^2}{3}}\right].\label{E0}
	\end{equation}
Substituting the derived expression (\ref{E0}) into Eq. (\ref{dmdx_dim})
and integrating the latter, we can obtain the next term
in the expansion for $\eta(x)$ and so on.
	
\begin{figure}[htb]
		\begin{center}
			\includegraphics[width=14cm]{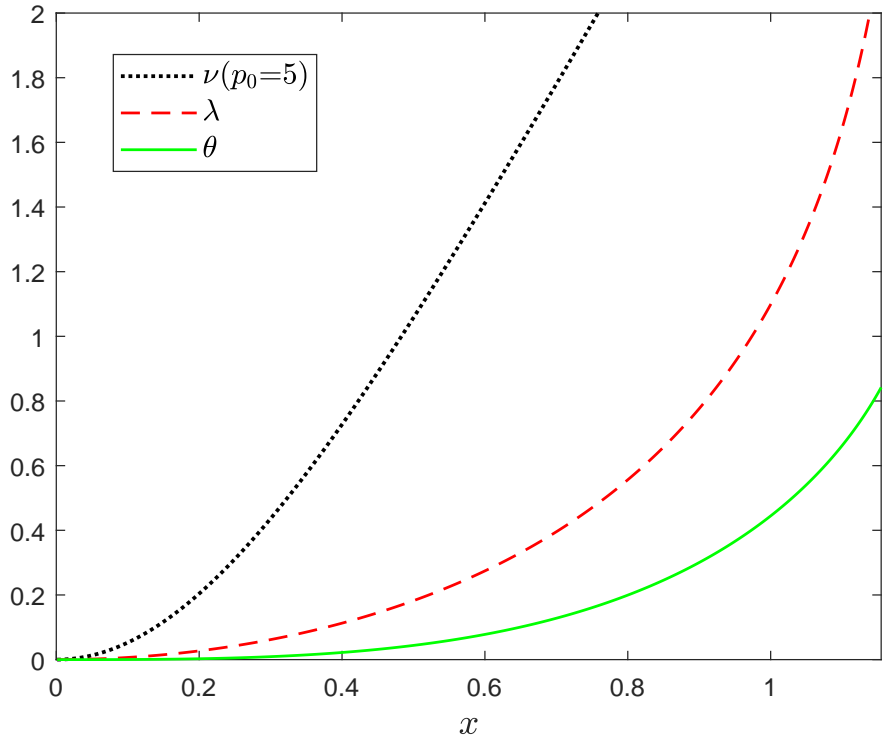}
		\end{center}
		\caption{Distributions of the metric functions $\lambda$ and $\nu$ at $p_0 {=}5$ and the dimensionless charge $\theta$  inside a star within $0\leq x \leq\frac{2}{\sqrt{3}}$.} \label{lambda_x}
	\end{figure}

Let us now turn to the function $\phi(x)=\sum\psi_i(x) Y_i$,
where we used the standard notation $Y_i\equiv n_i/\WI{n}{b}$. At
the boundary of the star $\phi=1$, while at its center
(see Eq. \ref{Phi_dim}) $\phi(0)\equiv\phi_0=1/\mathbf{c}(0)>1$. Given that,
according to (\ref{dCdr_dim}), $\mathbf{c}$ is a constant with an accuracy of
the order of $O(\WI{\Lambda}{q})$, Eq. (\ref{Phi_dim}) can be integrated with
the same accuracy as
	\begin{equation}
	\phi^{-1}(x)=\frac{3}{2}-\Bigg[\frac{3}{2}-\frac{1}{\phi_0}\Bigg]\sqrt{1{-}\frac{2x^2}{3}}.\label{Phi_solut}
	\end{equation}
It corresponds to the well-known solution (see, e.g.,
Synge 1960) for the pressure inside a homogeneous
incompressible fluid in general relativity (recall that in
dimensionless units $p=\phi-1$):
	\begin{equation}
	p(x)=\frac{(1{+}3p_0)\sqrt{1{-}\frac{2x^2}{3}}-(1{+}p_0)}{3(1{+}p_0)-(1{+}3p_0)\sqrt{1{-}\frac{2x^2}{3}}},\label{p_rigid}
	\end{equation}
where $p_0=p(0)$. To calculate a correction of the order
of $O(\WI{\Lambda}{q})$, it will suffice to integrate Eq. (\ref{dCdr_dim}) by setting
$\mathbf{c}(x)=\mathbf{c}(0)$ on the right-hand side and using Eq. (\ref{E0})
for $\theta(x)$. It is also easy to find the component $e^\nu=g_{00}$ of the metric in the first approximation. For our
case of $\rho=\mathrm{const}$, it is simply related to the pressure
(Weinberg 1972):
	\begin{equation}
	e^{\nu(x)}=\left[\frac{1+p_0}{1+p(x)}\right]^{2}.\label{nu_x}
	\end{equation}
The $\lambda$, $\theta$, and $\nu$ distributions for $p_0=5$ inside a star
are illustrated in Fig.~\ref{lambda_x}, which shows, in particular,
the importance of general relativity effects in the problem
under consideration.

\begin{figure}[htb]
		\begin{center}
			\includegraphics[width=14cm]{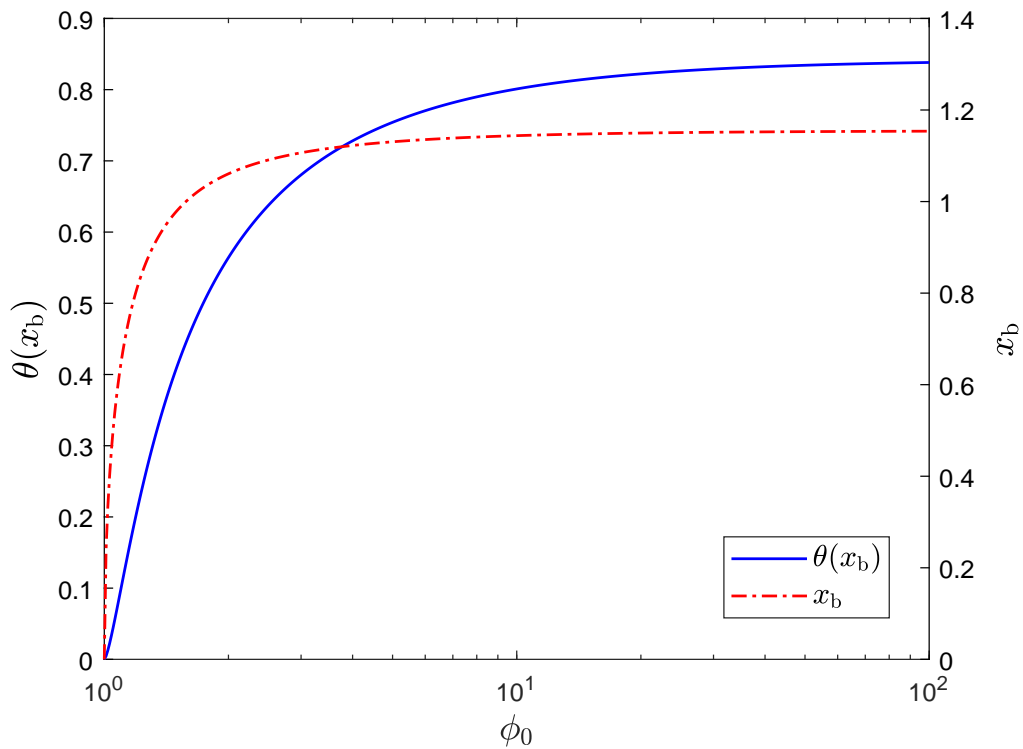}
		\end{center}
		\caption{Total dimensionless stellar charge $\theta(\WI{x}{b})$ and stellar boundary coordinate $\WI{x}{b}$ as a function of parameter $\phi_0\approx\WI{\psi}{n0}$.} \label{theta_xb}
	\end{figure}

The stellar boundary coordinate $\WI{x}{b}$ is found from
the condition $\phi(\WI{x}{b})=1$, which, according to (\ref{Phi_solut}),
gives, with an accuracy of the order of $O(\WI{\Lambda}{q})$,
	\begin{equation}
	\WI{x}{b}=\frac{\sqrt{6\big(\phi_0{-1}\big)\big(2\phi_0{-}1\big)}}{3\phi_0{-}2}.\label{x_b}
	\end{equation}
Obviously, $0\leq\WI{x}{b}\leq\frac{2}{\sqrt{3}}$. The quantities $\WI{x}{b}$ and
$\theta(\WI{x}{b})$ are shown in Fig.~\ref{theta_xb} as a function of $\phi_0$. The
maximum values of $\theta(\WI{x}{b}) \approx 0.84$ and $\WI{x}{b} \approx 1.15$ are
reached when $\phi_0\rightarrow \infty$.

Let us now turn to the equilibrium equations for
the matter components. Equation (\ref{Pi_dim}) for $\Pi(x)$ can
be integrated explicitly:
	\begin{equation}
	\Pi(x)=\ln\Bigg[\frac{3\phi_0{-}(3\phi_0{-}2)\sqrt{1{-}\frac{2x^2}{3}}}{2\sqrt{1{-}\frac{2x^2}{3}}}\Bigg].
	\end{equation}
Let us write Eq. (\ref{mu_equil_dim}) for the neutron chemical potential
(given that $\WI{\tilde{q}}{n}=0$):
	\begin{equation}
	\WI{\psi}{n}(x)=e^{\lambda(x)/2{-}\Pi(x)}\WI{\psi}{n}(0)=
	\frac{2\WI{\psi}{n}(0)}{3\phi_0{-}(3\phi_0{-}2)\sqrt{1{-}\frac{2x^2}{3}}}.\label{mu_n_sol}
	\end{equation}
Let us now consider $\phi(x)$:
	\begin{equation}
	\phi(x)=\sum_i \psi_i Y_i\approx
	\WI{\psi}{n}\WI{Y}{n}+(\WI{\psi}{p}{+}\WI{\psi}{e})\WI{Y}{p}=\WI{\psi}{n}(x),\label{Phi_mu_n}
	\end{equation}
where we first used the LEN condition (approximate,
with an accuracy of the order of $O(\WI{\Lambda}{q})$) and then the
beta-equilibrium condition. Expression (\ref{mu_n_sol}) for $\WI{\psi}{n}$
can then be rewritten with the same accuracy as
	\begin{equation}
	\WI{\psi}{n}(x)=
	\frac{2\phi_0}{3\phi_0{-}(3\phi_0{-}2)\sqrt{1{-}\frac{2x^2}{3}}}.\label{mu_n_sol_v2}
	\end{equation}
This expression is identically satisfied at $x = 0$
(given (\ref{Phi_mu_n})), while at $x=\WI{x}{b}$ defined by equality (\ref{x_b})
it leads to the limit $\WI{\psi}{n}(\WI{x}{b})=1$, as it must be.
	
\begin{figure}[htb]
		\begin{center}
			\includegraphics[width=14cm]{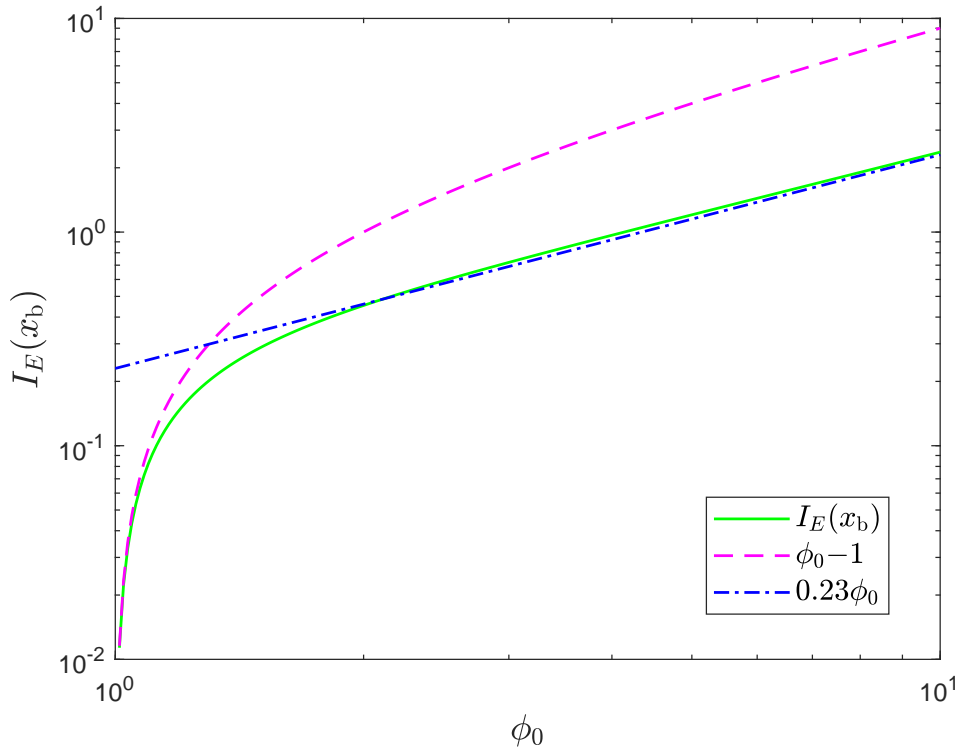}
		\end{center}
		\caption{Integral $\WI{I}{E}(\WI{x}{b})$ and its asymptotics when $\phi_0\rightarrow 1$ and $\phi_0\rightarrow \infty$.} \label{I_xb}
	\end{figure}

	Let us now turn to the case of charged particles.
We need to calculate the integral of the polarization
field in (\ref{mu_equil_dim}). It is
	\begin{multline}
	\WI{I}{E}(x)\equiv\int\limits_0^x\frac{\theta(x')}{x'^2}
	e^{\Pi(x')}dx'=\frac{3}{8}\left[\Big(3\phi_0-2-3\phi_0
	\sqrt{1{-}\frac{2x^2}{3}}\Big)\frac{1}{x}\sqrt{\frac{3}{2}}\arcsin{\sqrt{\frac{2}{3}}x}-\right.\\
	\left.{-}3\phi_0+4+(3\phi_0{-}2)\sqrt{1{-}\frac{2x^2}{3}}\right].\label{E_int}
	\end{multline}
The chemical potentials of the charged matter components are then
	\begin{equation}
	\WI{\psi}{p,e}(x)=\frac{\WI{\psi}{p,e}(0)\pm\WI{\alpha}{e}\WI{I}{E}(x)}{\frac{3}{2}\phi_0{-}
		\big(\frac{3}{2}\phi_0{-}1\big)\sqrt{1{-}\frac{2x^2}{3}}},\label{mu_pe}
	\end{equation}
where the upper and lower signs refer to the protons
and electrons, respectively. So far the parameter $\WI{\alpha}{e}$
has remained undetermined. Using (\ref{mu_pe}) and the
boundary condition for electrons $\WI{\psi}{e}(\WI{x}{b})=0$, we will
obtain
	\begin{equation}
	\WI{\alpha}{e}=\frac{\WI{\psi}{e}(0)}{\WI{I}{E}(\WI{x}{b})}.\label{alpha_def}
	\end{equation}
The integral $\WI{I}{E}(\WI{x}{b})$ here is a function of $\phi_0\approx\WI{\psi}{n}(0)\equiv\WI{\psi}{n0}$:
	\begin{equation}
	\WI{I}{E}(\WI{x}{b})=\frac{3}{8}\left[\frac{3\phi_0^2{-}6\phi_0{+}2}
	{\sqrt{(\phi_0{-}1)(2\phi_0{-}1)}}\arcsin\!\!\Big(\frac{2\sqrt{(\phi_0{-}1)
			(2\phi_0{-}1)}}{3\phi_0{-}2}\Big)
	{-}2(\phi_0{-}2)\right].
	\end{equation}
Its behavior is presented Fig.~\ref{I_xb}, where the following asymptotics are also shown:
	\begin{equation}
	\WI{I}{E}(\WI{x}{b}) =
	\begin{cases}
	\phi_0{-}1, & \phi_0\rightarrow 1,\\
	\frac{3}{4}\left[\frac{3}{2\sqrt{2}}\arcsin\Big(\frac{2\sqrt{2}}{3}\Big){-}1\right]\phi_0 \approx 0.23\phi_0, & \phi_0\rightarrow \infty.
	\end{cases}
	\end{equation}
In the non-relativistic limit, $\phi_0{-}1=\WI{\psi}{n0}{-}1$ is nothing
but the neutron Fermi energy $\WI{E}{Fe}^\mathrm{n}(0)$, which,
in view of the beta-equilibrium conditions, is $\WI{E}{Fe}^\mathrm{n}=\WI{E}{Fe}^{\mathrm{p}}+\WI{E}{Fe}^{\mathrm{e}}$. Given also that $\WI{\psi}{e0}=\WI{E}{Fe}^\mathrm{e}(0)$ for electrons,
we will obtain Eq. (\ref{alpha_def}) in the non-relativistic
case in the form
	\begin{equation}
	\WI{\alpha}{e}=\frac{\WI{E}{Fe}^\mathrm{e}(0)}{\WI{E}{Fe}^\mathrm{e}(0){+}\WI{E}{Fe}^\mathrm{p}(0)},
	\end{equation}
i.e., the old result for polytropes (see Eq. (II.22) in
Krivoruchenko et al. 2018).

	\section*{NUMERICAL EXAMPLES}
	\noindent To illustrate the results obtained, it is necessary
to understand the meaning of the parameter $\WI{\alpha}{e}$. For
this purpose, we will write the expression for the total stellar charge $Q$ within a sphere of radius $r$ as
	\begin{equation}
	Q(r)=4\pi r_0^3\WI{n}{b} \mathbf{e}\frac{\WI{\alpha}{e}}{\WI{\Lambda}{G}}\theta(x).\label{Q_tot}
	\end{equation}
	The number of baryons within the same sphere is
	\begin{equation}
	N(r)=4\pi r_0^3\WI{n}{b}\int e^{\lambda(x)/2}x^2 dx=4\pi r_0^3\WI{n}{b}\theta(x).\label{N_bar}
	\end{equation}
	The uncompensated charge per baryon in the star
is then given by the expression (see also Hund and
Kiessling 2021a, 2021b)
	\begin{equation}
	\frac{Q(r)}{N(r)}= \mathbf{e}\frac{\WI{\alpha}{e}}{\WI{\Lambda}{G}}.\label{QN}
	\end{equation}
In particular, the total stellar charge per baryon is
equal to the value of (\ref{QN}) at $x=\WI{x}{b}$.

At a characteristic number of baryons in a star
$\WI{N}{b}\sim 10^{57}$, an elementary charge $\mathbf{e}\approx 1.6\times 10^{-19}$~C,
and $\WI{\Lambda}{G}\sim 10^{36}$, we obtain a typical total stellar charge
$Q_{\mathrm{s}}\sim 100\WI{\alpha}{e}$~C (see also Krivoruchenko et al. 2018).
Thus, the parameter $\WI{\alpha}{e}\sim O(1)$ determines the total
stellar charge. As can be seen from Eq. (\ref{alpha_def}), it is
related to the electron chemical potential at zero. In
an ordinary NS under the assumption of LEN (and,
as a consequence, its zero total charge), specifying
one parameter (for example, the central density)
uniquely determines all properties of the configuration,
in particular, its mass. In our case, apart from
the parameter $\phi_0\approx\WI{\psi}{n0}$, we need to specify $\WI{\psi}{e0}$ ($\WI{\psi}{p0}$ is determined from the beta-equilibrium conditions),
thereby specifying both the total mass of the star and
its charge.
	
\begin{figure}[htb]
		\begin{center}
			\includegraphics[width=14cm]{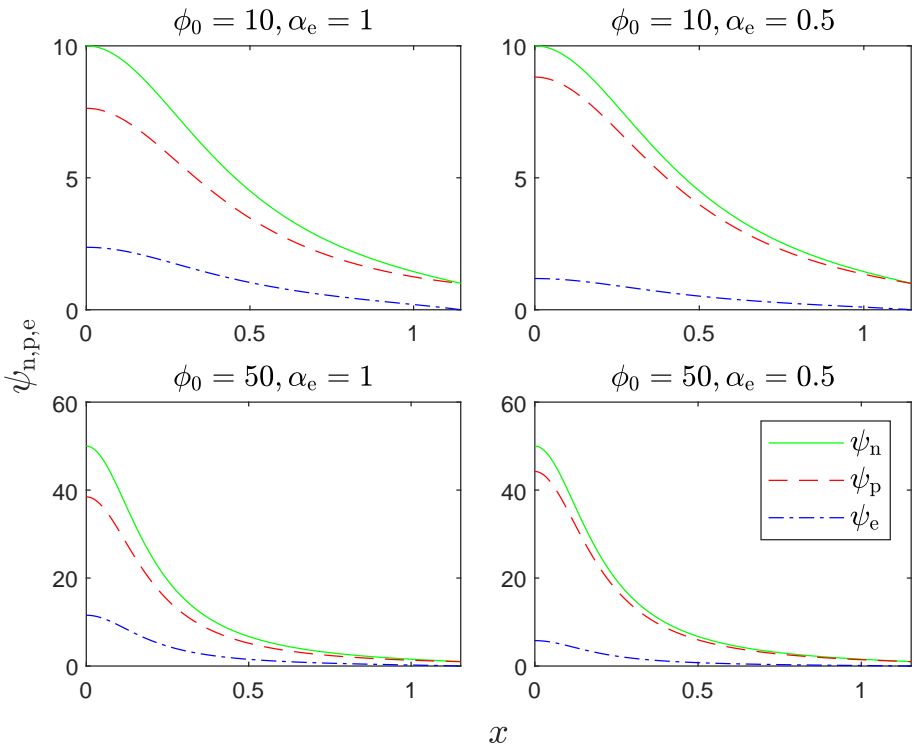}
		\end{center}
		\caption{Distributions of the dimensionless chemical potentials of neutrons, protons, and electrons inside a NS for several values
of $\phi_0$ and $\WI{\alpha}{e}$.} \label{mu_npe}
	\end{figure}

Several examples of the derived distributions of
the chemical potentials of the neutron, proton, and
electron components for various initial conditions are
shown in Fig.~\ref{mu_npe}. As $\WI{\alpha}{e}$ decreases, the electron chemical
potential also drops (see Eq. (\ref{alpha_def})), while $\WI{\psi}{p}$ approaches
$\WI{\psi}{n}$  (we can compare Eqs. (\ref{mu_n_sol_v2}) and (\ref{mu_pe}) and
take into account the beta-equilibrium condition).
	
\begin{figure}[htb]
		\begin{center}
			\includegraphics[width=14cm]{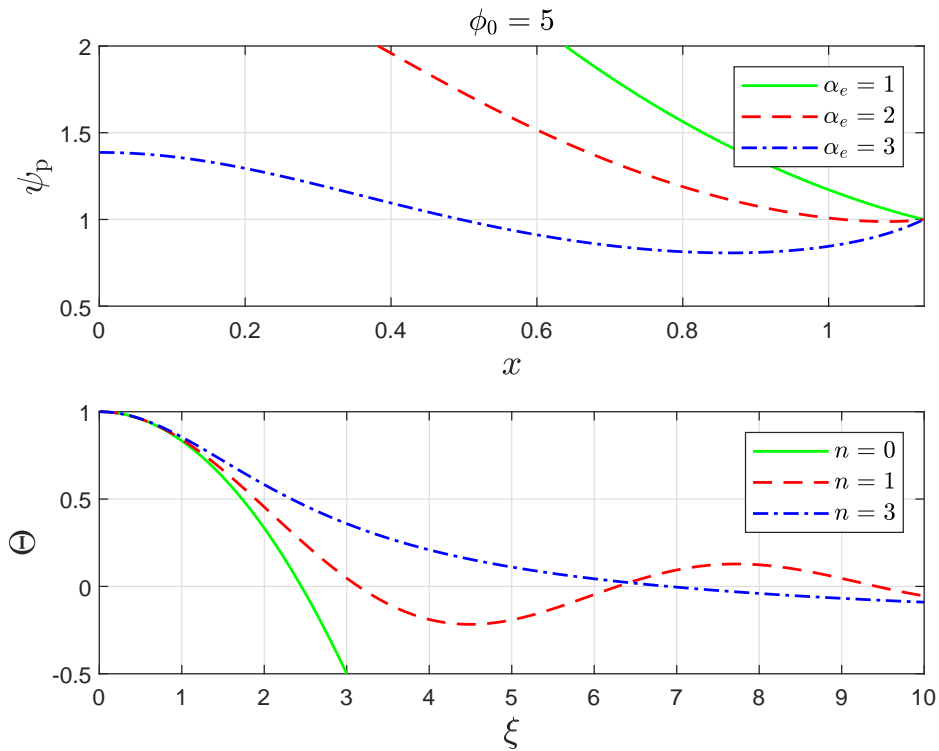}
		\end{center}
		\caption{Top: The behavior of the proton chemical potential near $\WI{\psi}{p}\sim 1$ for several values of $\WI{\alpha}{e}$. Bottom: The Emden functions
for several values of the polytropic index $n$.} \label{fig.mu_p}
	\end{figure}

How great can the parameter $\WI{\alpha}{e}$ and, hence, the
stellar charge be? It turns out that $\WI{\alpha}{e}$ much greater
than unity leads to an incorrect solution, as demonstrated
by Fig.~\ref{fig.mu_p}. The upper panel in the figure
shows the behavior of the proton chemical potential
near $\WI{\psi}{p}\sim 1$ for several values of $\WI{\alpha}{e}$ and fixed $\phi_0=5$.
At $\WI{\alpha}{e}=1$ the curve $\WI{\psi}{p}(x)$ with a finite slope runs
into unity at the boundary of the star. At $\WI{\alpha}{e}=2$ 
the slope at the boundary is almost zero, while at
$\WI{\alpha}{e}=3$ the curve runs into unity from below, from the
inadmissible (in our approximation) region $\WI{\psi}{p}<1$!
Thus, the solution for the proton chemical potential
is incorrect here, and the proton component, in fact,
ends at $x\approx 0.3$. This effect has a direct analog in
the theory of polytropes. The Edmen functions shown
on the lower panel of Fig.~\ref{fig.mu_p} for several values of the
polytropic index $n$ can also have several roots. For
example, the Edmen function for $n = 1$ is $\sin(\xi)/\xi$ and
becomes zero not only at $\xi=\pi$, but also at $\xi=2\pi$
etc. However, the solution lying at $\xi>\pi$ is already
unrealistic. At $\WI{\alpha}{e}\gtrsim 2$, although the solution obtained
formally satisfies the boundary condition, it is not
admissible in our case as well. Thus, large values
of $\WI{\alpha}{e}$ and, hence, the total stellar charge are also
inadmissible.

	%*************************************************************
	\section*{ELECTROSPHERE}
	\noindent Is it possible to obtain a globally electroneutral
stellar configuration with a total charge $Q_{\mathrm{s}}=0$ in the
approximation under consideration? Formula (\ref{QN})
seems to suggest that this is possible only at $\WI{\alpha}{e}=0$, i.e., in the case of strict local electroneutrality. However, this is not the case: everywhere above,
we used the significant condition that the number
densities of all matter components become zero \emph{at
one point}. If we relax this requirement, then it is
quite possible to construct a configuration in which
the number density of, for example, electrons inside
the star is lower than that of protons, but the electron
component itself extends slightly farther, forming
the so-called electrosphere on the stellar surface
(Krivoruchenko et al. 2018), thereby compensating
for the accumulated positive stellar charge. Let us
estimate the parameters of this electrosphere in our
approximation.
 
The electrosphere exists in a thin subsurface layer
of thickness $\WI{r}{e}$, with $\WI{r}{e} \ll \WI{r}{b}$, where the latter is
the radius of the boundary of the baryon component.
Hence, Eq. (\ref{dEdr}) for the charge $Q$ in the electrosphere
is easily integrated. The requirement of $Q_{\mathrm{s}}=0$ gives
the thickness of the electrosphere (cf. Eq. (A.9) from
Krivoruchenko et al. 2018):
\begin{equation}
	\frac{\WI{r}{e}}{\WI{r}{b}}=\frac{\WI{Q}{b}}{4\pi\WI{r}{b}^3 \mathbf{e}\WI{n}{e}}=\frac{\WI{\alpha}{e}}{\WI{\Lambda}{G}}\,\frac{\theta(\WI{x}{b})}{\WI{x}{b}^3 e^{\WI{\lambda}{b}/2}}\Big(\frac{\WI{n}{b}}{\WI{n}{e}}\Big) \sim 10^{-36},\label{r_e}
	\end{equation}
where $\WI{Q}{b}$ is the (positive) stellar charge at the boundary
of the baryon component. According to the equilibrium
equation (\ref{dmudr_new}), the electron chemical potential
inside this electrosphere is zero with an accuracy of
the order of $O(\WI{\Lambda}{G}^{-1})$.
	
	\section*{CONCLUSIONS}
	\noindent We considered the structure of a locally non-electroneutral
self-gravitating incompressible three-component
fluid in general relativity. As in the
case of Newtonian gravity, with regard to thermodynamic
quantities like the pressure $P$, the thermodynamic
potential  $\Phi$, etc. (see Eqs. (\ref{Phi_dim}), (\ref{dCdr_dim}) and (\ref{TOV_dim})), the solution is a series in the small parameter
$1/\WI{\Lambda}{G}\sim 10^{-36}$, where the first approximation
is the classical electroneutral solution. The solution
is degenerate, because it has no irregular component,
in complete agreement with the analogous
case in Newtonian gravity (see the Appendix from
Krivoruchenko et al. (2018)). We found that the nonelectroneutrality
may be neglected when calculating
such macroscopic NS parameters as the mass and
radius. Thus, we showed that the strange results
of Belvedere et al. (2012) cannot be simply a consequence
of LEN violation in general relativity. They
are probably associated with their interpretation of
the subtle effects occurring at the boundary of the
stellar core. An electrosphere can appear here in
our solution, but nothing like the sharp growth of
the electric field found by these authors occurs there.
The importance of a careful calculation of the phase
boundaries when solving the problems of plasma
polarization in astrophysical objects was pointed out
by Iosilevskiy (see, e.g., Iosilevskiy 2009).

The main feature to which the deviation from LEN
leads is contained in the equilibrium equations for the
chemical potentials of the matter components (\ref{mu_equil_dim}). In
them (for the charged components) the factor in front
of the integral consists of two multipliers compensating
for each other: one $\WI{\Lambda}{G}$ is huge, while the other
$(\WI{n}{p}{-}\WI{n}{e})/\WI{n}{b}$ is small. This factor denoted by  $\WI{\alpha}{e}=O(1)$
 enters into the final equilibrium equations for the
individual matter components (\ref{mu_pe}) and is responsible
for the (already large) effect of deviation from LEN.
In the long run, it also determines the total stellar
charge  (\ref{Q_tot}). This nuance is a characteristic and very
important feature of the problems of deviation from
LEN in stars: despite the fact that this effect (and the
polarization field generated by it) may be neglected
with regard to large-scale parameters, when passing
to the microlevel, it turns out that the force acting on
a charged particle from this field is comparable to the
gravitational force.

Despite the model nature of the problem considered,
its solution seems an important step on the path
of analyzing the structure of NSs with more realistic
equations of state in the absence of strict LEN.
	
	\vspace{1cm}
	The work of N.I. Kramarev was supported by the
``BAZIS'' Foundation for the Development of Theoretical
Physics and Mathematics (project no. 20-2-1-19-1). 

We are also grateful to the anonymous referees
whose remarks allowed our paper to be improved
significantly.
	
	\pagebreak
	%****************************************************************

	%xxxxxxxxxxxxxxxxxxxxxxxxxxxxxxxxxxxxxxxxxxxxxxxxxxxxxxxxxxxxx
	%---------------------------------------------------------------
\end{document}